\documentclass[pdflatex,sn-mathphys-num]{sn-jnl}

\usepackage{graphicx}%
\usepackage{multirow}%
\usepackage{amsmath,amssymb,amsfonts}%
\usepackage{amsthm}%
\usepackage{mathrsfs}%
\usepackage[title]{appendix}%
\usepackage{xcolor}%
\usepackage{textcomp}%
\usepackage{manyfoot}%
\usepackage{booktabs}%
\usepackage{algorithm}%
\usepackage{algorithmicx}%
\usepackage{algpseudocode}%
\usepackage{listings}%
\usepackage{ulem}%

\theoremstyle{thmstyleone}%
\theoremstyle{thmstyletwo}%

\theoremstyle{thmstylethree}%

\begin{document}

\title[Article Title]{Complex nematic dynamics driven by steady {Kolmogorov} flows}




\author*[1]{\fnm{Rahil N} \sur{Valani}}\email{rahil.valani@physics.ox.ac.uk}

\author[1,2]{\fnm{Sumesh P} \sur{Thampi}}\email{sumesh@iitm.ac.in}

\author[1]{\fnm{Julia M} \sur{Yeomans}}\email{julia.yeomans@physics.ox.ac.uk}

\affil*[1]{\orgdiv{Rudolf Peierls Centre for Theoretical Physics}, \orgname{University of Oxford}, \city{University of Oxford,
Oxford}, \postcode{OX1 3PU}, \country{United Kingdom}}

\affil[2]{\orgdiv{Department of Chemical Engineering}, \orgname{Indian Institute of Technology Madras}, \city{Chennai}, \postcode{600036}, \country{India}}




\abstract{Complex nematohydrodynamics, including irregular defect dynamics, is commonly associated with time-dependent flows, most notably in active turbulence. {Here we consider a simple model of a planar nematic subjected to a temporally steady but spatially varying flow and investigate the onset of time-dependent dynamics.} Subjecting a passive nematic to steady Kolmogorov flow, while eliminating velocity fluctuations and hydrodynamic feedback, we find that increasing the flow strength drives a transition from stationary to periodic and, at stronger forcing, irregular spatiotemporal dynamics. The time-dependent states involve repeated local suppression of nematic order and director reorganisation, arising from the competition between spatially varying shear and nematic relaxation. Bifurcation and phase-space analyses characterise this progression, while reduced-order models identify {the} dynamical origin{s}. A two-mode model captures the onset of periodic motion through a saddle-node infinite-period bifurcation. A five-mode Galerkin truncation further recovers the progression from stationary to periodic and irregular dynamics. These results {allow us to explain how} a fixed, spatially heterogeneous velocity field can generate temporal complexity through the nonlinear dynamics of nematic order alone.}




\maketitle


\section{Introduction}

Nematic order arising from the local alignment of anisotropic
constituents occurs in systems ranging from molecular and
colloidal liquid crystals to cytoskeletal suspensions and
cellular tissues
~\cite{de1993physics,lewis2014colloidal,
doostmohammadi2022physics,maroudas2021topological}.
Topological defects, where orientational order breaks down,
naturally arise in these systems and can play an important role
in {determining} their dynamics. In biological systems, defects have been
associated with processes including cell extrusion and death
~\cite{saw2017topological}, morphogenesis
~\cite{maroudas2021topological}, and the organisation of
collective cellular motion
~\cite{doostmohammadi2022physics,Shankar2022}. They therefore
provide not only signatures of broken orientational order, but
also localised structures around which stresses, flows and
biological processes can be organised.

A striking example of such dynamics occurs in active nematics.
Internally generated active stresses destabilise orientational
order and produce active turbulence, characterised by the
continuous creation, annihilation and motion of oppositely
charged topological defects
~\cite{giomi2015geometry,Doostmohammadi2018_ActiveNematics,
alert2022active}. In particular, $+1/2$ defects exhibit
characteristic self-propelled motion, while the interaction of
$\pm1/2$ defects with the surrounding fluctuating flow
contributes to the irregular spatiotemporal dynamics of active
turbulence
~\cite{giomi2015geometry,Doostmohammadi2018_ActiveNematics,
Thampi2016}.

{Rich nonequilibrium dynamics can also arise in passive
nematics subjected to external driving. Electrical, thermal
and mechanical forcing can distort nematic order, induce
transitions between distinct director configurations and
generate topological defects
~\cite{dubois1971hydrodynamic,manneville1981transition,
weiss2013nematic,Jewell2009,Sengupta2013,
aplinc2016porous,giomi2017cross,Copar2020-qu,
mur2022continuous}. In our recent work on two-dimensional
pressure-driven channel flows of passive nematics, increasing
the pressure gradient produced periodic topological
transitions followed by defect-mediated spatiotemporal chaos
~\citep{Valani_Thampi_Yeomans_2026}. As in active turbulence,
however, the velocity field in this system is itself
dynamical. Local velocity gradients reorient the nematic,
while changes in nematic order feed back onto the flow through
elastic stresses. The resulting dynamics therefore reflect
the coupled evolution of the nematic and flow, making it
difficult to separate the response of the nematic to spatial
velocity gradients from the effects of temporal variations in
the flow.}

Here we isolate these effects in a substantially reduced
setting. We restrict the nematic to planar order and prescribe
a spatially varying velocity field that remains strictly
steady in time. The velocity has no dynamical
degrees of freedom: there are no velocity fluctuations,
backflow feedback or hydrodynamic instabilities, and only the
nematic tensor evolves. Despite this reduction, increasing the
flow strength drives the nematic from a stationary state to
periodic dynamics and, at stronger forcing, to irregular
spatiotemporal behaviour. This provides a controlled setting
in which the temporal complexity arises entirely from the
nonlinear response of the nematic to a fixed, heterogeneous
velocity field.

{In two dimensions, the
nematic order is described by the symmetric, traceless tensor
\begin{equation}
\mathbf{Q}
=
\begin{pmatrix}
Q_{xx} & Q_{xy}\\
Q_{xy} & -Q_{xx}
\end{pmatrix}
=
\frac{S}{2}
\begin{pmatrix}
\cos 2\theta & \sin 2\theta\\
\sin 2\theta & -\cos 2\theta
\end{pmatrix},
\label{eq:Qdef}
\end{equation}
which contains only two local degrees of freedom, the
magnitude of nematic order $S$ and the director orientation
$\theta$~\cite{Beris1994,giomi2015geometry,Shankar2022,
D3SM01684F}. This restriction is important in relation to the
earlier steady-shear literature. Chaotic dynamics in
spatially homogeneous and spatially varying alignment-tensor
models have generally been obtained with additional
three-dimensional orientational degrees of freedom
~\cite{Grosso2001,Forest2004,RienaeckerKroegerHess2002,
Chakrabarti2004}. In-plane models under steady Couette shear
have also exhibited spatially heterogeneous periodic states,
including coexisting tumbling and wagging regions accompanied
by local loss of orientational order
~\cite{Forest2005,Forest2008}.} {These studies establish that
steady shear can support a wide range of nematic dynamics.
Here we use a strictly planar two-component tensor and a
simple{,} spatially varying steady flow as a minimal setting to ask {the} question: how do the dynamical
transitions between stationary, periodic and more complex
states arise{, and how many interacting nematic modes are
required to describe them?}}


{The $Q$-tensor description is particularly suited to {address} this
question because the magnitude of nematic order $S$ can
evolve and locally vanish~\cite{Beris1994}. Flow-induced
rotation, elastic relaxation and relaxation of the magnitude
of nematic order nonlinearly couple the components of
$\mathbf Q$, while spatial variations introduce interactions
between different nematic modes. The resulting system is
therefore sufficiently simple to permit systematic
reduced-order modelling, while retaining the nonlinear
ingredients required for transitions between distinct
dynamical states.}


{Here {s}tarting from the
Beris--Edwards equations, we prescribe a temporally steady
Kolmogorov flow and characterise the progression from
stationary to periodic and irregular dynamics. We then use
Galerkin projection to identify the minimal modal interactions
underlying these transitions. A two-mode model captures the
bifurcation to periodic motion, while a five-mode model
reveals how additional spatial degrees of freedom enable more
complex dynamics.}

The remainder of the paper is organised as follows.
In Sec.~\ref{sec:model}, we introduce the nematohydrodynamic
model, reduce it to a unidirectional flow with transverse
spatial variations, and formulate the prescribed steady-flow
problem. In Sec.~\ref{sec:full_dynamics}, we examine the
resulting nematic dynamics as the imposed flow strength is
varied and characterise the associated bifurcation structure.
Motivated by these results, in Secs.~\ref{sec:two_mode} and
\ref{sec:five_mode} we construct two- and five-mode Galerkin
reductions to examine the origin of the periodic and irregular
dynamics. In Sec.~\ref{sec:phase_diagram}, we map the
different dynamical regimes as functions of flow strength and
flow alignment. Finally, in Sec.~\ref{sec:discussion}, we
discuss the broader implications {of our analysis} for driven and active
nematics.


\section{Nematohydrodynamic model}\label{sec:model}

\subsection{Order-parameter dynamics}

We consider a two-dimensional nematic described by the symmetric, traceless order parameter tensor $\mathbf{Q}$ defined in Eq.~(\ref{eq:Qdef}).
Its dynamics are governed by the Beris--Edwards equation
\begin{equation}
\partial_t Q_{ij}
+
u_k\partial_k Q_{ij}
=
\mathcal{S}_{ij}
+
\Gamma H_{ij},
\label{eq:BE}
\end{equation}
where $\Gamma$ is the inverse relaxation time and $H_{ij}$ is the molecular field. The co-rotational flow-alignment term is
\begin{align}
\mathcal{S}_{ij}
={}&
(\lambda E_{ik}+\Omega_{ik})
\left(Q_{kj}+\frac{\delta_{kj}}{2}\right)
\nonumber+
\left(Q_{ik}+\frac{\delta_{ik}}{2}\right)
(\lambda E_{kj}-\Omega_{kj})
\nonumber\\
&-
2\lambda
\left(Q_{ij}+\frac{\delta_{ij}}{2}\right)
Q_{kl}\partial_k u_l ,
\label{eq:Sij}
\end{align}
where
\begin{equation*}
E_{ij}
=
\frac{1}{2}
(\partial_i u_j+\partial_j u_i),
\qquad
\Omega_{ij}
=
\frac{1}{2}
(\partial_j u_i-\partial_i u_j),
\end{equation*}
are the strain-rate and vorticity tensors, respectively, and $\lambda$ is the flow alignment parameter.

The molecular field is described by
\begin{equation}
H_{ij}
=
\chi Q_{ij}
\left(
S_{\rm nem}^2-S^2
\right)
+
K\nabla^2Q_{ij},
\label{eq:H}
\end{equation}
where
\begin{equation}
S^2
=
4\left(Q_{xx}^2+Q_{xy}^2\right).
\label{eq:S}
\end{equation}
The first term relaxes the magnitude of nematic order towards its preferred value $S_{\rm nem}$, with $\chi$ setting the relaxation strength, while the elastic constant $K$ penalises spatial variations of ${Q}_{ij}$.

Equations~(\ref{eq:BE})--(\ref{eq:S}) describe the dynamics of the orientational order parameter tensor $\mathbf{Q}$ when subjected to a flow field but without any hydrodynamic feedback. The flow gradients rotate and align ${Q}_{ij}$ through
Eq.~(\ref{eq:Sij}), while the molecular field drives relaxation of the field through Eq.~(\ref{eq:H}).




\subsection{Unidirectional reduction for flow field}

We {first} restrict the system to a unidirectional flow with
spatial variations only in the transverse direction,
\begin{equation*}
\mathbf{u}=u(y,t)\hat{\mathbf{x}},
\qquad
\mathbf{Q}=\mathbf{Q}(y,t).
\end{equation*}
Since all fields are independent of $x$, the advective term
$u_k\partial_k Q_{ij}$ vanishes. Defining the local shear rate
\begin{equation*}
\dot{\gamma}(y,t)=\partial_yu,
\end{equation*}
Eq.~(\ref{eq:BE}) reduces to
\begin{align}
\partial_tQ_{xx}
={}&
\Gamma K\partial_y^2Q_{xx}
+
Q_{xy}\dot{\gamma}
+
\Gamma \chi Q_{xx}
\left(S_{\rm nem}^2-S^2\right)
\nonumber\\
&-
2\lambda Q_{xx}Q_{xy}\dot{\gamma},
\label{eq:qxx1D}
\\[2mm]
\partial_tQ_{xy}
={}&
\Gamma K\partial_y^2Q_{xy}
-
Q_{xx}\dot{\gamma}
+
\Gamma \chi Q_{xy}
\left(S_{\rm nem}^2-S^2\right)
\nonumber\\
&+
\frac{\lambda}{2}\dot{\gamma}
-
2\lambda Q_{xy}^2\dot{\gamma}.
\label{eq:qxy1D}
\end{align}
The flow coupling is therefore explicit: it is not the velocity, but the  shear rate $\dot{\gamma}${,} that drives the evolution of $\mathbf{Q}$ through Eqs.~(\ref{eq:qxx1D}) and (\ref{eq:qxy1D}).

%


\subsection{Nematic dynamics in a steady flow}

We now impose a steady Kolmogorov flow,
\begin{equation}
u(y)=U_0\sin(k_0y),
\qquad
k_0=\frac{2\pi}{L_0} .
\label{eq:fixedflow}
\end{equation}
The nematic is consequently driven by the stationary shear
profile
\begin{equation}
\dot{\gamma}(y)
=
U_0k_0\cos(k_0y),
\label{eq:shear}
\end{equation}
whose amplitude is controlled by $U_0$. Thus the imposed shear contains a single spatial Fourier mode,
which will later provide a natural basis for constructing
low-dimensional Galerkin reductions of the nematic dynamics.

The resulting dynamical system consists solely of
Eqs.~(\ref{eq:qxx1D}) and (\ref{eq:qxy1D}), with the shear
field in Eq.~(\ref{eq:shear}) imposed externally. The velocity
has no dynamical degrees of freedom and there is no feedback
from $\mathbf{Q}$ onto the flow. Nevertheless, the two
components of the order tensor remain nonlinearly coupled
through the flow-alignment terms and through the relaxation of
their combined magnitude (Eq.~\eqref{eq:H}-\eqref{eq:S}). 
Equivalently, these terms couple the magnitude $S$ and
orientation $\theta$ of the nematic order. 

Mathematically, Eqs.~(\ref{eq:qxx1D}) and
(\ref{eq:qxy1D}) may also be viewed as a two-component,
spatially forced reaction--diffusion system, in which nematic
elasticity provides diffusive coupling and the local shear and
bulk relaxation provide the coupled reaction dynamics
~\cite{CrossHohenberg1993,CrossGreenside2009}.

This formalism separates the temporal complexity of the nematic from that of the surrounding flow. 
Any stationary, periodic or irregular dynamics that emerge in $\mathbf{Q}$ under Eq.~(\ref{eq:shear}) arise from a strictly time independent, driving velocity field. 
The prescribed model therefore provides a minimal setting to test whether complex nematic dynamics can originate within the nematic order parameter {$\mathbf{Q}$} itself.


\subsection{Numerical methods}
\label{sec:numerics}

We solve Eqs.~(\ref{eq:qxx1D}) and (\ref{eq:qxy1D}) using the
method of lines. The periodic domain $-L/2\leq y\leq L/2$ is
discretised using $N=100$ grid points, with centred finite
differences for the spatial derivatives. The resulting $2N$
coupled ordinary differential equations are integrated using
MATLAB's adaptive explicit Runge--Kutta solver
\texttt{ode45}, with relative and absolute tolerances of
$10^{-7}$ and $10^{-8}$, respectively. 
 Unless otherwise
stated, we use
\begin{equation*}
L=L_0=20,\quad
\Gamma=1,\quad
K=0.01,\quad
\chi=0.1,\quad
\lambda=0,\quad
S_{\rm nem}=1.
\end{equation*}
The prescribed velocity and shear are evaluated analytically
from Eqs.~(\ref{eq:fixedflow}) and (\ref{eq:shear}).

The initial condition is a weakly perturbed aligned nematic,
\begin{equation*}
Q_{xx}(y,0)
=
\frac{S_0}{2}\cos[2\theta_0(y)],
\quad
Q_{xy}(y,0)
=
\frac{S_0}{2}\sin[2\theta_0(y)],
\end{equation*}
where $S_0=1$ and
$\theta_0(y)=0.1[2\pi r(y)-\pi]$, with $r$ uniformly
distributed on $[0,1]$. 

For the bifurcation analysis, each trajectory is evolved to
$t=5000$ with an output interval of $\Delta t=0.2$, and the
first $60\%$ of the trajectory is discarded as transient.
Representative phase-space trajectories are also integrated to $t=10000$, with the portion $t\geq5000$ used to determine the long-time dynamics.

The states emerging from the long-time dynamics are characterised by the local maxima
and minima in time of the spatially averaged nematic order
\begin{equation}
\langle S\rangle(t)
=
\frac{1}{L}
\int_{-L/2}^{L/2}S(y,t)\,{\rm d}y .
\label{eq:Saverage}
\end{equation}
A state is classified as steady when the late-time spread of $\langle S\rangle$ is smaller than $10^{-2}$. For
time-dependent states, periodicity is determined from the
recurrence of successive extrema of $\langle S\rangle$.
A state is classified as periodic if its late-time extrema
repeat within a prescribed tolerance after at most eight
oscillation cycles, thereby capturing limit cycles up to
period eight. States showing no such recurrence are classified
as irregular. 



\begin{figure*}
    \centering
    \includegraphics[width=0.95\textwidth]{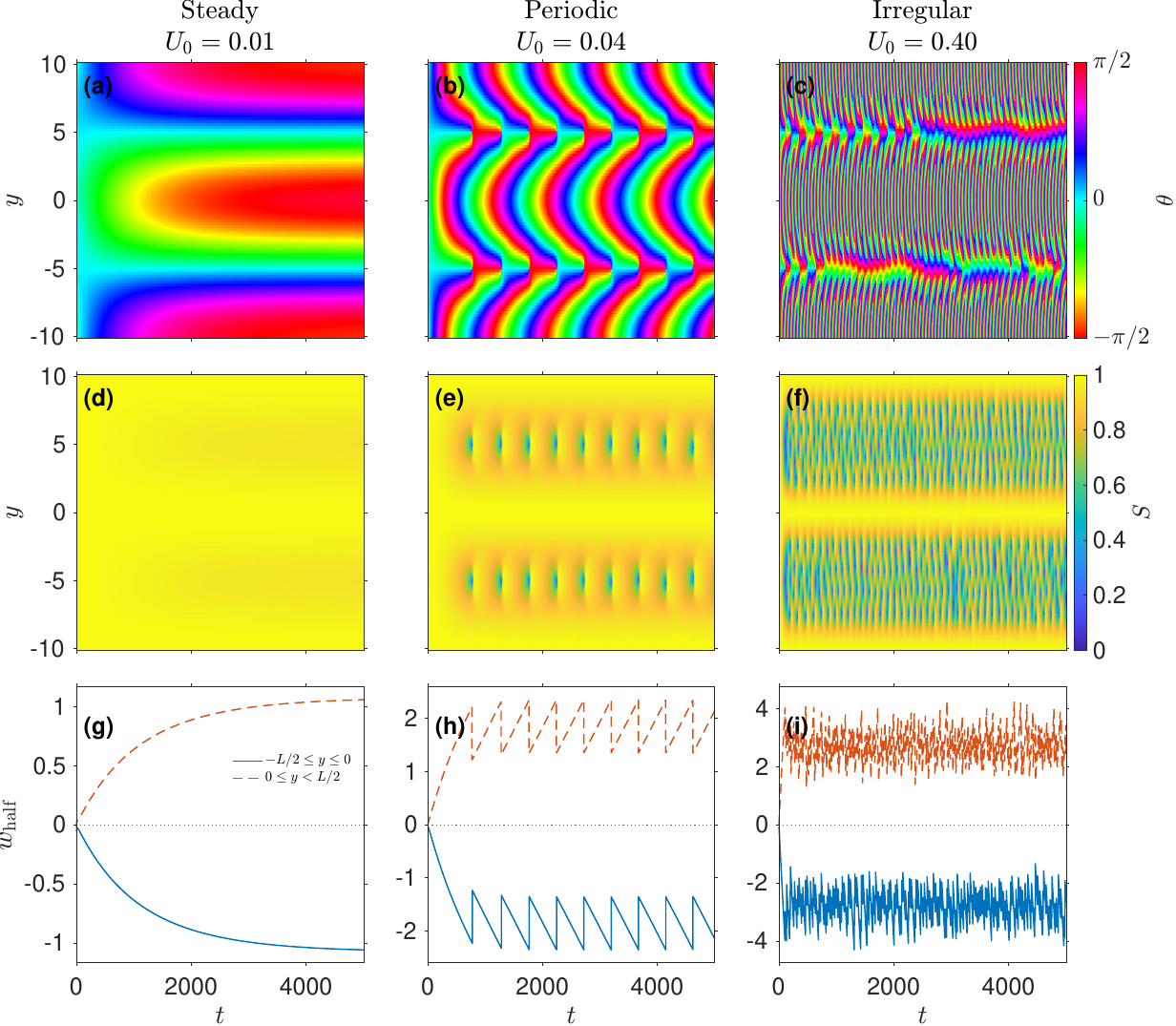}
  \caption{
\textbf{Complex nematic dynamics.}
Kymographs of the director orientation $\theta(y,t)$
(top row) and scalar nematic order
$S(y,t)$ (middle row), together
with the accumulated nematic rotations $w_-$ and $w_+$ over
$-L/2\leq y\leq0$ and $0\leq y<L/2$, respectively
(bottom row), under the prescribed steady Kolmogorov flow
$u_x(y)=U_0\sin(k_0y)$.
For $U_0=0.01$, the nematic relaxes to a stationary
configuration, with equal and opposite director rotations
across the two halves of the domain [(a),(d),(g)].
For $U_0=0.04$, periodic director reorientations are
accompanied by repeated local suppression of nematic order
near $y\simeq\pm L/4$ and regular discontinuous changes in
$w_-$ and $w_+$ of opposite sign [(b),(e),(h)]; see also Supplemental Movie 1.
At stronger forcing, $U_0=0.40$, these low-order events and
the corresponding changes in $w_-$ and $w_+$ become irregular
[(c),(f),(i)].
In all cases the velocity field remains fixed in time and only
$Q_{xx}$ and $Q_{xy}$ evolve.
}
    \label{fig:steady_periodic_irregular}
\end{figure*}

\begin{figure*}
    \centering
    \includegraphics[width=\textwidth]{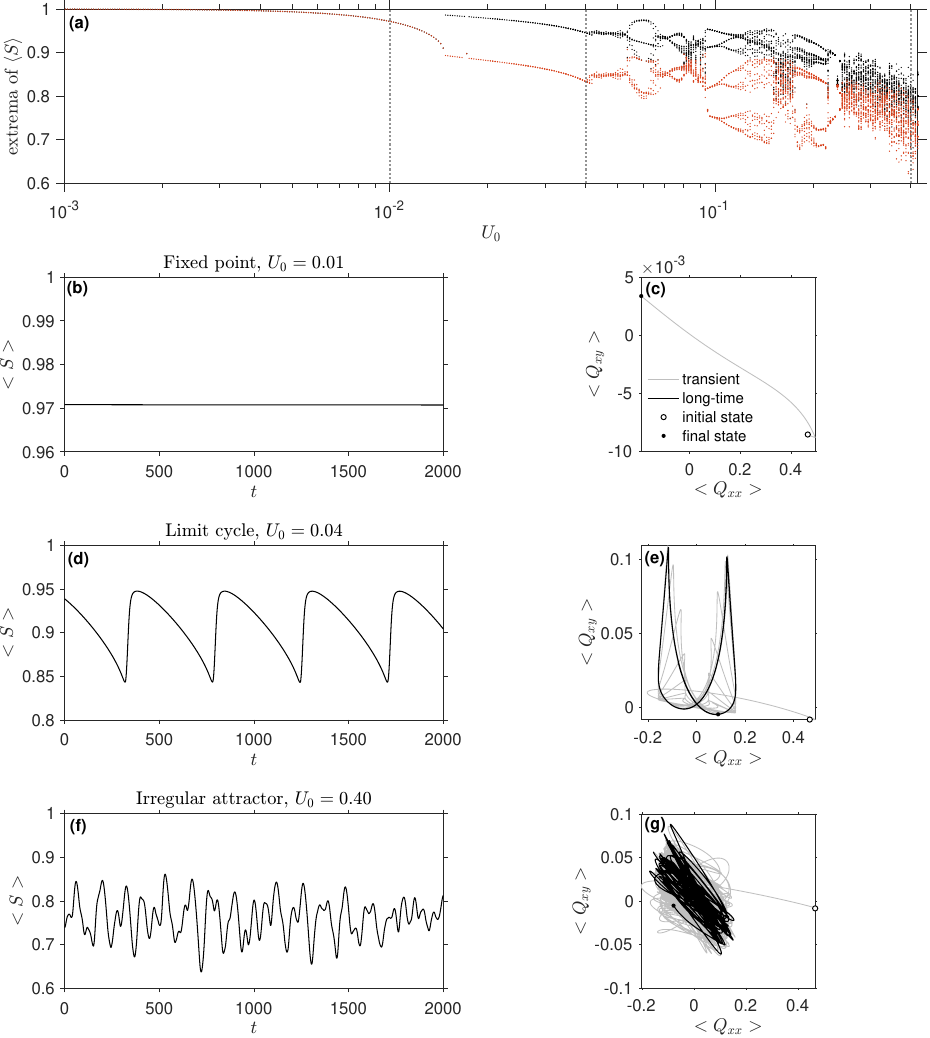}
    \caption{
    \textbf{Transition from steady to complex nematic dynamics under steady flow.}
    (a) Bifurcation diagram constructed from the local maxima (black)
    and minima (red) of the spatially averaged nematic order
    $\langle S\rangle$ as the amplitude $U_0$
    of the prescribed Kolmogorov flow is increased.
    Vertical dotted lines indicate the three representative states
    shown in the remaining panels.
    (b,d,f) Long-time evolution of $\langle S\rangle$ for
    $U_0=0.01$, $0.04$, and $0.40$, respectively.
    At weak forcing the nematic approaches a stationary state (b),
    intermediate forcing produces sustained periodic oscillations (d),
    while stronger forcing generates irregular dynamics with
    non-repeating variations in the magnitude of nematic order (f).
    (c,e,g) Corresponding trajectories projected onto the
    $(\langle Q_{xx}\rangle,\langle Q_{xy}\rangle)$ plane.
    Transient trajectories are shown in grey and the long-time
    dynamics in black; open and filled circles denote the initial and
    final states, respectively.
    The projections illustrate the evolution from a fixed point (c)
    to a closed limit cycle (e) and, at strong forcing, to an
    irregular attractor (g).
    }
    \label{fig:bifurcation_phase_space}
\end{figure*}

\section{Complex nematic dynamics under steady flow}
\label{sec:full_dynamics}

{Figure~\ref{fig:steady_periodic_irregular} reveals {the}
progression of nematic dynamics under a simple, temporally
steady flow, from a stationary texture to sustained periodic
motion and, at stronger forcing, irregular spatiotemporal
dynamics.} We fix all parameters and vary only the
amplitude $U_0$ of the imposed Kolmogorov flow. In addition to
the director orientation $\theta$ and scalar order $S$, we
monitor the accumulated nematic rotation over each half of the
domain,
\begin{align}
w_-(t)
&=
\frac{1}{2\pi}
\int_{-L/2}^{0}
\partial_y(2\theta)\,{\rm d}y,
\\
w_+(t)
&=
\frac{1}{2\pi}
\int_{0}^{L/2}
\partial_y(2\theta)\,{\rm d}y .
\label{eq:half_winding}
\end{align}
%
{The spatial structure of the Kolmogorov flow provides a simple
physical picture for the dynamics. The imposed velocity
$u_x(y)=U_0\sin(k_0y)$ has extrema at $y=\pm L/4$, where the
local shear rate
$\dot{\gamma}(y)=U_0k_0\cos(k_0y)$ vanishes. The director at these points therefore experiences no direct
shear-induced rotation. Immediately on either side, however,
the shear rate has opposite signs and drives neighbouring
directors to rotate in opposite senses. This tends to build up
a local orientational mismatch around $y=\pm L/4$, while
elasticity acts to smooth the resulting distortion. The
response is therefore controlled by the competition between
flow-induced differential rotation and elastic and nematic
relaxation. As the forcing is increased, this competition
gives rise to the distinct stationary, periodic and irregular
regimes described below.}

At weak forcing, $U_0=0.01$, the nematic accommodates this
spatially varying shear through a stationary distortion
[Fig.~\ref{fig:steady_periodic_irregular}(a,d,g)]. In this
regime, elastic relaxation is sufficient to balance the
shear-induced director distortion, preventing the build-up of
strong orientational gradients. The magnitude of nematic order
therefore remains close to its equilibrium value, while the
director develops equal and opposite accumulated rotations
across the two halves of the domain. After the initial
relaxation, both $\theta(y)$ and $S(y)$ are independent of
time and $w_\pm$ approach constant values.

A qualitatively different response appears at intermediate
forcing, $U_0=0.04$
[Fig.~\ref{fig:steady_periodic_irregular}(b,e,h)]. The
stronger shear drives the neighbouring directors on either
side of the zero-shear points in opposite senses, building up
orientational distortion faster than it can be relaxed by
elasticity. Once this distortion becomes sufficiently strong,
the nematic order is locally suppressed near
$y\simeq\pm L/4$. The director is then weakly defined in these
low-order regions and can rapidly reorganise, releasing the
accumulated distortion. This reorganisation appears as abrupt,
oppositely signed changes in $w_-$ and $w_+$ in
Fig.~\ref{fig:steady_periodic_irregular}(h). Nematic order
subsequently recovers, after which the steady shear begins to
build up the distortion again. The dynamics therefore form a
repeating cycle of distortion, local loss of order, director
reorganisation and recovery, providing a simple physical
picture for the sustained periodic motion. This process is illustrated in Supplementary
Movie~1, which shows the evolving director texture, nematic
order and accumulated rotations $w_\pm$ together with the
imposed Kolmogorov flow profile.

These events also provide a natural connection to defect
dynamics. In two dimensions, the nematic order
vanishes at a defect core, allowing the director to reorganise
as the defect moves through the system. If such a defect were
to cross a line, an
observer restricted to that line would see a local suppression
of $S$ followed by an abrupt change in the director, similar
to the events observed here [see, for example,
Fig.~6 of \citep{Valani_Thampi_Yeomans_2026}]. Alternatively, a local suppression of nematic order followed by
director reorganisation may be viewed as the one-dimensional
signature of the creation of an oppositely charged defect
pair in two dimensions, with the two defects subsequently
moving away from the creation site in opposite directions. Thus, while the present
model does not contain defects themselves, it captures the
local loss of order and director reorganisation associated
with defect dynamics in higher dimensions. 

At stronger forcing, $U_0=0.40$
[Fig.~\ref{fig:steady_periodic_irregular}(c,f,i)], the same
basic mechanism persists but loses its temporal regularity.
Local suppression of nematic order occurs more frequently near
the zero-shear regions, allowing repeated director
reorganisations that no longer follow a simple periodic
sequence. Correspondingly, $w_-$ and $w_+$ exhibit frequent
irregular changes, in contrast to the regular repeating events
of the periodic state. The nematic thus develops strongly
irregular spatiotemporal dynamics despite the imposed velocity
field remaining strictly steady in time.





Having established the progression from stationary to periodic
and irregular nematic dynamics, we next characterise the
transitions between these regimes. Figure~\ref{fig:bifurcation_phase_space}(a)
shows the local maxima and minima of the spatially averaged
nematic order $\langle S\rangle(t)$ as a function of $U_0$.
The variations in $\langle S\rangle$ capture the repeated
suppression and recovery of nematic order associated with the
director reorganisations in
Fig.~\ref{fig:steady_periodic_irregular}, providing a compact
measure of the underlying dynamical transitions.


At small $U_0$, the maxima and minima coincide, reflecting
relaxation to a stationary state. With increasing forcing they
separate, signalling the onset of sustained oscillations in the
nematic order. With further increase in $U_0$, the bifurcation diagram
develops multiple branches and windows of more complex
behaviour. At strong forcing, the extrema populate broad bands
of values, indicating the loss of simple periodicity and the
onset of irregular time dynamics.

The representative time series in
Fig.~\ref{fig:bifurcation_phase_space}(b,d,f) make this
transition particularly transparent. For $U_0=0.01$,
$\langle S\rangle$ approaches a constant. At $U_0=0.04$, $\langle S\rangle$ executes a stable periodic oscillation, with each cycle associated with the repeated suppression and recovery of nematic order and the accompanying winding-number transitions shown in Fig.~\ref{fig:steady_periodic_irregular}. At $U_0=0.40$, by contrast, both the amplitudes and intervals between successive oscillations vary irregularly.

A complementary view is obtained by projecting the dynamics
onto the spatially averaged tensor components,
\begin{equation*}
\langle Q_{ij}\rangle
=
\frac{1}{L}
\int_{-L/2}^{L/2}Q_{ij}(y,t)\,{\rm d}y.
\end{equation*}
The corresponding trajectories in the
$(\langle Q_{xx}\rangle,\langle Q_{xy}\rangle)$ plane are
shown in Fig.~\ref{fig:bifurcation_phase_space}(c,e,g).
At weak forcing the trajectory converges to a fixed point,
whereas at intermediate forcing it approaches a closed orbit,
consistent with a stable limit cycle. At strong forcing the
trajectory instead explores an extended region of the
projected phase space without settling onto a single closed
orbit.

Together, these measures illustrate a transition from a stationary
fixed point to periodic motion and, at stronger forcing, to
irregular long-time dynamics. Since the imposed flow remains
steady, these transitions arise from the dynamics of the
nematic field itself. We therefore first focus on the onset of
periodic motion and ask whether this transition can be
understood from a minimal set of nematic modes.

\begin{figure*}
    \centering
    \includegraphics[width=\textwidth]{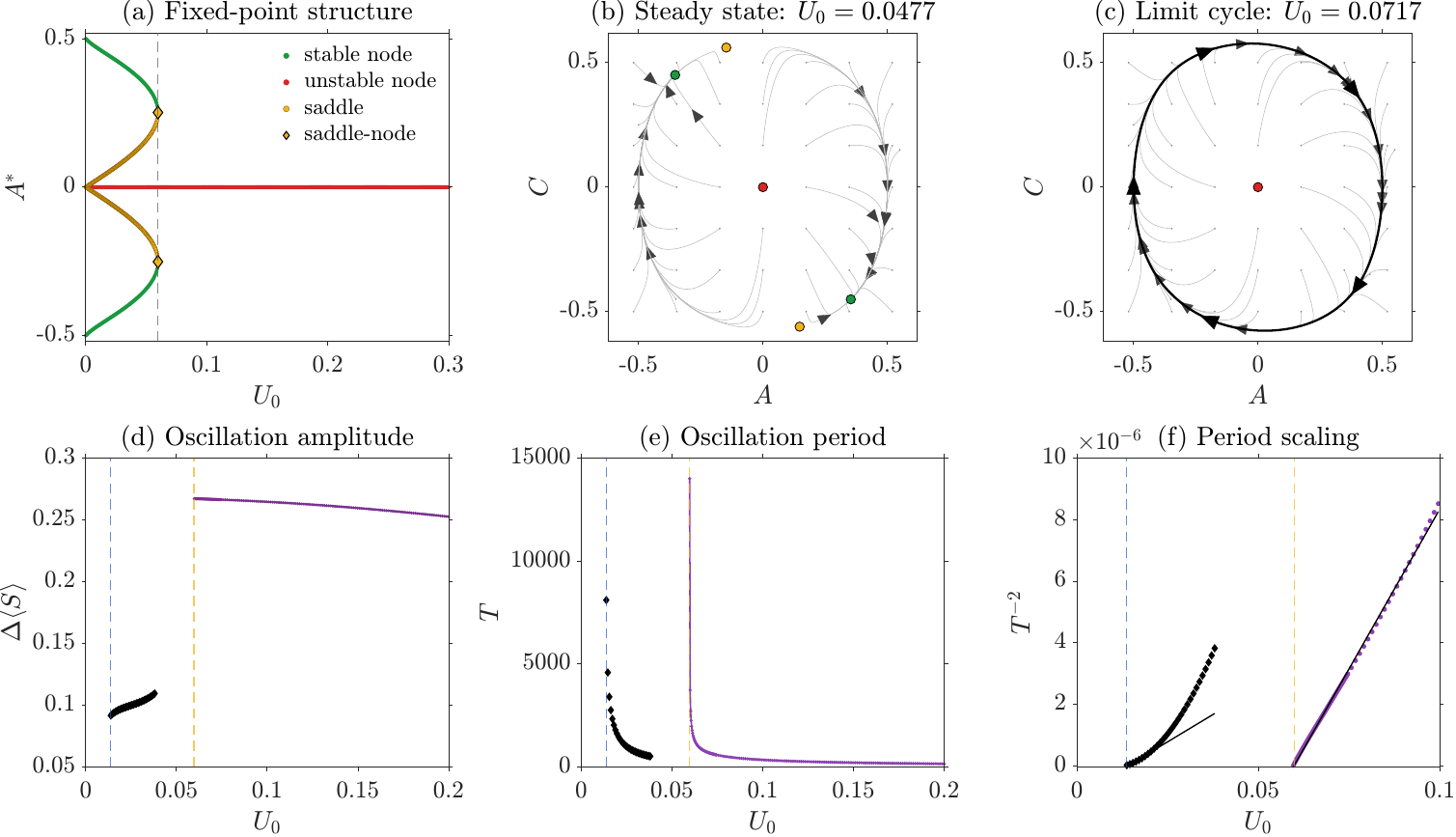}
  \caption{
\textbf{Two-mode model and comparison with the onset of periodic
dynamics in the full PDE.}
(a) Fixed points of the two-mode system as a function of the
imposed-flow amplitude $U_0$. Stable nodes, unstable nodes and
saddles are shown in green, red and yellow, respectively. The
stable and saddle branches collide at saddle-node bifurcations
(diamonds; dashed vertical line).
(b) Phase portrait below the bifurcation
($U_0=0.0477$). Grey curves show trajectories from different
initial conditions approaching the two stable
fixed points {(green)}. Yellow points denote saddles and the origin is an
unstable fixed point.
(c) Above the saddle-node ($U_0=0.0717$), the stable fixed
points and saddles have disappeared and trajectories approach
the stable limit cycle shown in black. Arrows indicate the
direction of motion.
(d) Oscillation amplitude $\Delta\langle S\rangle$ and
(e) oscillation period $T$ for the two-mode model (violet) and the full
spatially resolved PDE (black). The critical forcing differs between
the two systems (blue and yellow dashed lines), but in both cases periodic motion appears with
finite amplitude and the period increases strongly towards
onset. Dashed vertical lines indicate the corresponding
transition values.
(f) Near-onset scaling of $T^{-2}$. The approximately linear
behaviour close to the transition is consistent with the
infinite-period scaling of a saddle-node infinite-period
(SNIPER) bifurcation. The two-mode model therefore captures
the main features of the steady-to-periodic transition observed
in the full PDE, despite the shift in its critical forcing.
}
    \label{fig:two_mode_SNIPER}
\end{figure*}


\section{Reduced model for the onset of periodic dynamics}
\label{sec:two_mode}

{Here, w}e construct the simplest Galerkin truncation capable of
describing the transition from the stationary state to
 periodic motion observed in the full system. 
%
Since the imposed shear contains the fundamental mode
$\cos(k_0y)$, we retain a uniform component of $Q_{xx}$ and
the leading shear-coupled harmonic of $Q_{xy}$, {of amplitude $C$,}
\begin{equation}
Q_{xx}(y,t)=A(t),
\qquad
Q_{xy}(y,t)=C(t)\cos(k_0y).
\label{eq:two_mode_ansatz}
\end{equation}
%
Substituting 
Eq.~(\ref{eq:two_mode_ansatz}) into Eq.~\eqref{eq:S},
\begin{equation*}
S^2
=
4\left[
A^2+C^2\cos^2(k_0y)
\right].
\end{equation*}
%
Defining
\begin{equation*}
\kappa=\Gamma K,
\qquad
\mu=\Gamma \chi,
\qquad
G=U_0k_0,
\end{equation*}
{and taking $\lambda=0$,}
Eqs.~(\ref{eq:qxx1D}) and (\ref{eq:qxy1D}) become
\begin{align}\label{eq:A_give}
\dot A
={}&
GC\cos^2(k_0y)
+
\mu A
\left[
S_{\rm nem}^2
-4A^2
-4C^2\cos^2(k_0y)
\right].
\end{align}
\begin{align}\label{eq:C_give}
\dot C\cos(k_0y)
={}&
-\kappa k_0^2C\cos(k_0y)
-GA\cos(k_0y)
\nonumber\\
&+
\mu C\cos(k_0y)
\left[
S_{\rm nem}^2
-4A^2
-4C^2\cos^2(k_0y)
\right].
\end{align}
Since $A$ is the spatially uniform mode, we project Eq.~\eqref{eq:A_give} onto the constant Fourier mode by averaging over one
spatial period. Equation~\eqref{eq:C_give} is projected onto the retained
$\cos(k_0y)$ mode by multiplying by $\cos(k_0y)$ and averaging
over one spatial period. 
{The} resulting two-dimensional autonomous system is {then}
\begin{align*}
\dot A
&=
\mu A
\left(
S_{\rm nem}^2-4A^2-2C^2
\right)
+
\frac{G}{2}C,
\\
\dot C
&=
-\kappa k_0^2C
+
\mu C
\left(
S_{\rm nem}^2-4A^2-3C^2
\right)
-
GA.
\end{align*}
The linear terms proportional to $G$ couple the two retained
modes through the imposed shear, while the cubic terms arise
from relaxation of the magnitude of nematic order. These
nonlinearities are sufficient for the reduced system to
support both stationary states and periodic motion.

Figure~\ref{fig:two_mode_SNIPER}(a) shows the fixed-point
structure of the reduced system. At weak forcing there are two
symmetry-related stable nodes, each separated from the unstable
origin by a saddle. The corresponding phase portrait in
Fig.~\ref{fig:two_mode_SNIPER}(b) shows trajectories
approaching one of the two stable stationary states. As $U_0$
is increased, each stable node approaches a neighbouring saddle
and the pairs annihilate in saddle-node bifurcations. Beyond
this point no stable stationary state remains and trajectories
approach a stable limit cycle
[Fig.~\ref{fig:two_mode_SNIPER}(c)].

The onset of oscillations in the two-mode model occurs through
a saddle-node infinite-period (SNIPER) bifurcation. The
oscillation amplitude $\Delta\langle S\rangle$, defined as the
difference between the maximum and minimum of $\langle S\rangle$
over one cycle, remains finite at the transition
[Fig.~\ref{fig:two_mode_SNIPER}(d)], while the period $T$
increases sharply on approaching the saddle-node
[Fig.~\ref{fig:two_mode_SNIPER}(e)]. For a SNIPER
bifurcation~\citep{Strogatz2015_NonlinearDynamicsChaos},
\begin{equation}
T\sim(U_0-U_c)^{-1/2},
\end{equation}
so that $T^{-2}$ varies linearly with $U_0-U_c$. The
near-linear scaling observed close to onset
[Fig.~\ref{fig:two_mode_SNIPER}(f)], together with the
saddle-node collision and the slow passage of the periodic
orbit through the former saddle-node region, confirms the
SNIPER character of the transition in the reduced model.

The full PDE shows qualitatively similar behaviour near the
onset of periodic motion. As shown in
Figs.~\ref{fig:two_mode_SNIPER}(d) and (e), oscillations emerge
with finite amplitude and their period increases strongly on
approaching the transition. Moreover, sufficiently close to
onset, $T^{-2}$ shows an approximately linear dependence on
$U_0$, although with a different slope from that of the
two-mode model [Fig.~\ref{fig:two_mode_SNIPER}(f)]. Further
from onset, the PDE data deviate from this linear scaling,
reflecting dynamics not captured by the two-mode truncation.
Thus, while the SNIPER bifurcation is established explicitly
only for the reduced model, the finite-amplitude onset,
increasing period and near-onset scaling of the full PDE are
consistent with a similar infinite-period mechanism.

The two-mode model therefore provides a minimal dynamical
description of the steady-to-periodic transition. The imposed
shear couples the amplitudes $A$ and $C$ and drives their
rotational dynamics, while bulk and elastic relaxation shape
the stationary states and their stability. Increasing the
forcing eliminates the stable stationary states through
saddle-node bifurcations, leaving a stable limit cycle whose
period becomes large near onset. Thus, the model identifies
the competition between shear-induced rotation and nematic
relaxation as a simple mechanism for the transition to
periodic motion. The irregular dynamics observed at stronger
forcing require additional spatial degrees of freedom,
motivating the higher-mode truncation considered next.

\begin{figure*}
    \centering
    \includegraphics[width=0.9\textwidth]
    {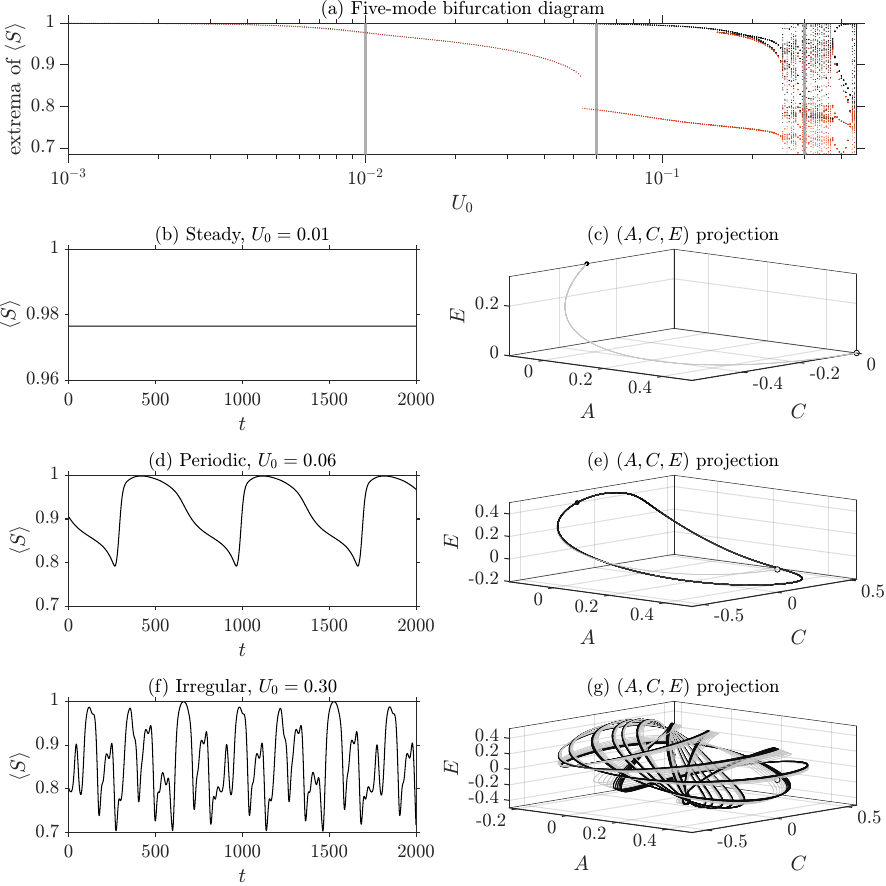}
    \caption{
    \textbf{Dynamics of the five-mode Galerkin model.}
    (a) Bifurcation diagram constructed from the local maxima
    (black) and minima (red) of the spatially averaged nematic
    order $\langle S\rangle$ as the imposed-flow amplitude
    $U_0$ is varied.
    At weak forcing the dynamics approach a stationary state,
    while increasing $U_0$ produces periodic oscillations and,
    at stronger forcing, {an} increasingly complex bifurcation
    structure.
    (b,d,f) Representative time series of $\langle S\rangle$
    for a steady state at $U_0=0.01$, a periodic state at
    $U_0=0.06$, and an irregular state at $U_0=0.30$,
    respectively.
    (c,e,g) Corresponding trajectories projected onto the
    $(A,C,E)$ subspace of the five modal amplitudes.
    Grey curves denote the transient and black curves the
    long-time dynamics.
    The five-mode truncation therefore retains the progression
    from stationary to periodic {to} irregular dynamics observed
    in the full spatially resolved model, while providing a
    low-dimensional representation of the underlying dynamics.
    }
    \label{fig:five_mode}
\end{figure*}


\section{Five-mode model and the emergence of irregular dynamics}
\label{sec:five_mode}
We now consider {a second} compact model, a five-mode truncation approximation that retains interactions between the leading shear-generated spatial harmonics. Guided by the spatial structure of the imposed shear, and for
$\lambda=0$, we retain three even harmonics of $Q_{xx}$ and two
odd harmonics of $Q_{xy}$,
\begin{align}
Q_{xx}(y,t)
&=
A(t)
+
B(t)\cos(2k_0y)
+
F(t)\cos(4k_0y),
\label{eq:five_qxx}
\\
Q_{xy}(y,t)
&=
C(t)\cos(k_0y)
+
E(t)\cos(3k_0y).
\label{eq:five_qxy}
\end{align}
The two-mode approximation is recovered by setting
$B=E=F=0$. 

Substituting Eqs.~(\ref{eq:five_qxx}) and
(\ref{eq:five_qxy}) into the governing Eqs.~\eqref{eq:qxx1D}-\eqref{eq:qxy1D}, and
projecting onto the retained Fourier modes as discussed in Sec.~\ref{sec:two_mode}, gives a
five-dimensional autonomous system of the form
\begin{equation}
\dot{\mathbf a}
=
\mathbf{F}(\mathbf a;U_0),
\qquad
\mathbf a=(A,B,C,E,F).
\label{eq:five_vector}
\end{equation}
The explicit equations and their derivation are given in
Appendix~\ref{app:five_mode_derivation}. The reduced system
retains shear-induced coupling between the tensor components,
nonlinear relaxation of the magnitude of nematic order, and
elastic damping of spatial variations.

Figure~\ref{fig:five_mode} shows the resulting dynamics. As
for the full PDE, the bifurcation diagram is constructed from
the long-time extrema of $\langle S\rangle$. At weak forcing,
the maxima and minima coincide, corresponding to a stationary
state [Fig.~\ref{fig:five_mode}(a)]. Increasing the forcing
produces separated extrema associated with periodic motion,
followed at stronger forcing by a more complex
distribution of extrema.

Representative trajectories shown in Figs.~\ref{fig:five_mode}(b)-(g) illustrate these regimes. At
$U_0=0.01$, $\langle S\rangle$ approaches a constant
[Fig.~\ref{fig:five_mode}(b)] and the trajectory converges to
a fixed point in the $(A,C,E)$ projection
[Fig.~\ref{fig:five_mode}(c)]. At $U_0=0.06$, sustained
oscillations of $\langle S\rangle$
[Fig.~\ref{fig:five_mode}(d)] correspond to a closed orbit in
the projected state space [Fig.~\ref{fig:five_mode}(e)].
Thus, the five-mode model retains the steady-to-periodic
transition already captured by the two-mode truncation.

At stronger forcing, the additional degrees of freedom allow
more complex dynamics. For $U_0=0.30$,
$\langle S\rangle$ exhibits irregular oscillations
[Fig.~\ref{fig:five_mode}(f)], while the corresponding
trajectory no longer approaches a simple closed orbit
[Fig.~\ref{fig:five_mode}(g)]. Instead, it explores an
extended region of the projected state space. The five-mode
system can therefore reproduce the qualitative progression
from fixed-point to periodic {to} irregular dynamics observed
in the full spatially resolved model.


\begin{figure*}
    \centering
    \includegraphics[width=\columnwidth]{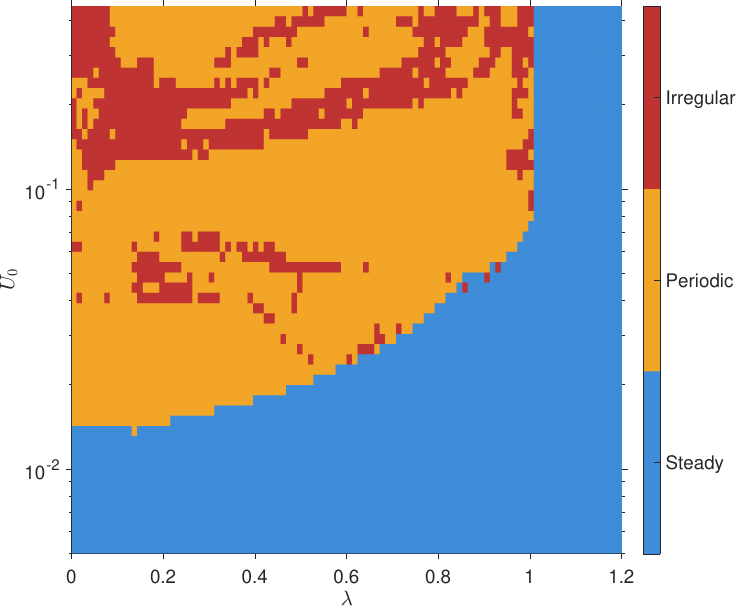}
    \caption{
    \textbf{Dynamical phase diagram under steady flow.}
    Long-time dynamics from solution of PDE{s} in the
    $(\lambda,U_0)$ plane at $100\times100$ parameter space resolution, where $\lambda$ is the flow-alignment
    parameter and $U_0$ is the amplitude of the prescribed, steady Kolmogorov flow
    $u_x(y)=U_0\sin(k_0y)$.
    The dynamical states are classified from the asymptotic extrema
    of the spatially averaged nematic order
    $\langle S\rangle$.
    Steady states (blue) exhibit negligible temporal variation of
    $\langle S\rangle$; periodic states (orange) are identified
    by a finite number {(less than or equal to 8)} of repeating extrema; and irregular states
    (red) exhibit a broad {range} {(more than 8)} of non-repeating extrema.
    }
    \label{fig:phase_diagram}
\end{figure*}


\section{Phase diagram}
\label{sec:phase_diagram}


The preceding results focused on $\lambda=0$. We now examine
how the long-time dynamics of the PDEs depend on both the flow
strength and the flow-alignment parameter.
Figure~\ref{fig:phase_diagram} shows the resulting dynamical
phase diagram in the $(\lambda,U_0)$ plane. As described in
Sec.~\ref{sec:numerics}, a state is classified as steady when
the late-time variation of $\langle S\rangle$ is negligible.
Time-dependent states are classified as periodic when their
extrema recur with period less than or equal to eight, and as irregular otherwise.

For $0\leq\lambda<1$, increasing $U_0$ generally drives the
system from steady to time-dependent dynamics. Periodic states
occupy a substantial region of the phase diagram, together
with irregular windows that become more prevalent at stronger
forcing. The non-monotonic boundaries and alternating
periodic and irregular regions indicate a sequence of
dynamical transitions as the flow strength is varied. The overall structure of the phase diagram is robust to spatial
resolution, although the detailed boundaries and interspersed
windows of periodic and irregular dynamics show some
resolution dependence (Appendix~\ref{app:grid_convergence}).

A marked change occurs near $\lambda=1$. For $\lambda>1$, all
states observed over the parameter range considered here are
steady. This change coincides with the
familiar distinction between flow-tumbling and flow-aligning
nematics. For a well-ordered nematic, a stationary director
orientation under simple shear becomes possible in the
flow-aligning regime, whereas the director continues to
reorient in the flow-tumbling regime
~\cite{de1993physics}. 

\section{Discussion and conclusion}
\label{sec:discussion}

Complex defect dynamics are most familiar in active nematics,
where active stresses generate flows that reorient the nematic
and drive defect creation, annihilation and motion
~\cite{giomi2015geometry,Doostmohammadi2018_ActiveNematics,
Thampi2016,alert2022active}. The orientational and
hydrodynamic dynamics are then closely coupled, making it
difficult to separate the response of the nematic to spatial
velocity gradients from that associated with temporal
fluctuations of the flow.

Here we separate these effects by prescribing a spatially
varying but temporally steady flow and removing hydrodynamic
feedback. Despite this reduction, the nematic progresses from
stationary to periodic and irregular dynamics as the forcing
is increased. In Appendix~\ref{app:backflow}, we relax this
assumption and allow the velocity field to evolve. Although backflow introduces
time-dependent corrections to the velocity field, the
qualitative nematic dynamics persist, supporting the use of
a prescribed steady flow model to isolate their underlying
mechanism. The physical origin of this behaviour can be
understood from the competition between flow-induced rotation
and nematic relaxation. The spatially varying shear drives
neighbouring regions of the nematic at different rotation
rates, while elasticity resists the resulting orientational
distortion. At intermediate forcing, distortion builds up
around the zero-shear regions until nematic order is locally
suppressed, allowing the director to reorganise and release
the accumulated distortion. Order then recovers and the
process repeats, producing periodic motion; at stronger
forcing, these reorganisation events become frequent and
irregular.

The reduced-order models provide a complementary dynamical
description of this mechanism. In the two-mode model,
shear-induced coupling and nematic relaxation produce a
saddle-node transition to periodic motion consistent with a
SNIPER bifurcation, while additional spatial harmonics in the
five-mode model allow irregular long-time dynamics to emerge.
These models demonstrate how increasingly complex dynamics
can arise from nonlinear interactions among only a few
nematic modes. Importantly, in the full system sustained
periodic and irregular dynamics are observed only for
$\lambda<1$, while the response is stationary for
$\lambda>1$ over the parameter range studied. This change
near $\lambda=1$ coincides with the familiar transition from
flow-tumbling to flow-aligning behaviour, emphasising that the
ability of the director to continuously reorient under shear
is closely connected to the emergence of the time-dependent
states.

The local loss of order accompanying these reorganisations
provides a connection to defect dynamics in two dimensions.
Although the present model cannot support isolated
$\pm1/2$ defects, it captures two key features of
defect-mediated dynamics: local suppression of nematic order
and rapid director reorganisation. The model therefore
provides a minimal setting for studying the local
order-parameter dynamics associated with defect-like events,
without explicitly resolving the defects themselves.

{A natural extension is to test this mechanism directly in two
dimensions, where the resulting creation, annihilation and
motion of $\pm1/2$ defects can be followed explicitly. This
may also provide insight into defect generation in active
turbulence. There, spatially heterogeneous velocity gradients
continually distort the director field, but evolve in time
together with the nematic. Freezing an instantaneous velocity
field from an active-turbulent state and subsequently evolving
only the nematic order parameter would test whether its
spatial structure alone can build sufficient orientational
distortion to suppress nematic order and nucleate defects.
Comparing this with the fully
time-dependent system would reveal how much of the defect
dynamics is driven by the spatial structure of the velocity
gradients and how much requires temporal flow fluctuations. The
strong dependence on flow alignment found here further
suggests that such a mechanism should be most effective in
the flow-tumbling regime, consistent with the sustained
time-dependent dynamics being restricted to $\lambda<1$ in
the present model.}

Our results therefore identify a simple physical route by
which a fixed, heterogeneous velocity field can
generate temporal complexity in a planar nematic. Steady shear
gradients first produce a stationary distorted texture; with
increasing forcing, the competition between differential
rotation and nematic relaxation generates repeated local loss
of order and periodic director reorganisation, followed at
stronger forcing by irregular dynamics. The reduced models
show that the onset of this behaviour can be understood from a
small number of interacting nematic modes. More broadly, these
results separate spatial complexity of the driving flow from
its temporal complexity and show that the former alone can
generate rich time-dependent nematic dynamics.


\section*{Acknowledgements}

R.V. acknowledges useful discussions with \href{https://www.maths.ox.ac.uk/people/ruth.baker} {Prof. Ruth Baker} (Mathematical Institute, University of Oxford). R.V. acknowledges the support of the Leverhulme Trust [Grant No. LIP-2020-014]. R.V. and J.M.Y. acknowledge the support of the ERC Advanced Grant ActBio (funded as UKRI Frontier Research Grant EP/Y033981/1). S.P.T. thanks the Royal Society and the Wolfson Foundation for the Royal Society Wolfson Fellowship award and acknowledges the support of the Department of Science and Technology, India via the research grant CRG/2023/000169.


\appendix

\section{Derivation of the five-mode Galerkin model}
\label{app:five_mode_derivation}

 We consider Eqs.~\eqref{eq:qxx1D}-\eqref{eq:qxy1D} with $\lambda=0$. 
Defining
\begin{equation}
\kappa=\Gamma K,
\qquad
\mu=\Gamma \chi,
\qquad
G=U_0k_0,
\qquad
x=k_0y,
\end{equation}
we use the five-mode ansatz
\begin{align}
Q_{xx}
&=
A+B\cos2x+F\cos4x,
\label{eq:app_ansatz_xx}
\\
Q_{xy}
&=
C\cos x+E\cos3x.
\label{eq:app_ansatz_xy}
\end{align}
The even--odd harmonic structure is preserved by the imposed
shear and nonlinear terms, while higher harmonics generated by
the nonlinearity are discarded by the Galerkin projection.

To evaluate the nonlinear terms, {we} write
\begin{equation}
R=Q_{xx}^2+Q_{xy}^2
=
r_0+r_2\cos2x+r_4\cos4x+r_6\cos6x+r_8\cos8x,
\end{equation}
where
\begin{align}
r_0
&=
A^2+\frac{B^2+F^2+C^2+E^2}{2},
\\
r_2
&=
2AB+BF+\frac{C^2}{2}+CE,
\\
r_4
&=
2AF+\frac{B^2}{2}+CE,
\\
r_6
&=
BF+\frac{E^2}{2},
\\
r_8
&=
\frac{F^2}{2}.
\end{align}
The retained Fourier coefficients of $Q_{xx}R$ and $Q_{xy}R$
are
\begin{align}
N_0
&=
A r_0+\frac{B r_2+F r_4}{2},
\\
N_2
&=
A r_2
+B\left(r_0+\frac{r_4}{2}\right)
+\frac{F}{2}\left(r_2+r_6\right),
\\
N_4
&=
A r_4
+\frac{B}{2}\left(r_2+r_6\right)
+F\left(r_0+\frac{r_8}{2}\right),
\\
M_1
&=
C\left(r_0+\frac{r_2}{2}\right)
+\frac{E}{2}\left(r_2+r_4\right),
\\
M_3
&=
E\left(r_0+\frac{r_6}{2}\right)
+\frac{C}{2}\left(r_2+r_4\right).
\end{align}

Projecting Eqs.~(\ref{eq:qxx1D}) and
(\ref{eq:qxy1D}) onto the retained Fourier modes using
their orthogonality then gives
\begin{align}
\dot A
={}&
\mu S_{\rm nem}^2 A
+\frac{G}{2}C
-4\mu N_0,
\label{eq:five_A}
\\
\dot B
={}&
\left(
\mu S_{\rm nem}^2-4\kappa k_0^2
\right)B
+\frac{G}{2}(C+E)
-4\mu N_2,
\label{eq:five_B}
\\
\dot C
={}&
\left(
\mu S_{\rm nem}^2-\kappa k_0^2
\right)C
-G\left(A+\frac{B}{2}\right)
-4\mu M_1,
\label{eq:five_C}
\\
\dot E
={}&
\left(
\mu S_{\rm nem}^2-9\kappa k_0^2
\right)E
-\frac{G}{2}(B+F)
-4\mu M_3,
\label{eq:five_E}
\\
\dot F
={}&
\left(
\mu S_{\rm nem}^2-16\kappa k_0^2
\right)F
+\frac{G}{2}E
-4\mu N_4.
\label{eq:five_F}
\end{align}
These equations form the closed five-dimensional system used
in Sec.~\ref{sec:five_mode}. The imposed shear couples
neighbouring even and odd harmonics, elasticity increasingly
damps the higher spatial modes, and the cubic bulk terms
couple the retained amplitudes through the local magnitude of
nematic order.

\begin{figure}[t]
    \centering
    \includegraphics[width=\linewidth]
    {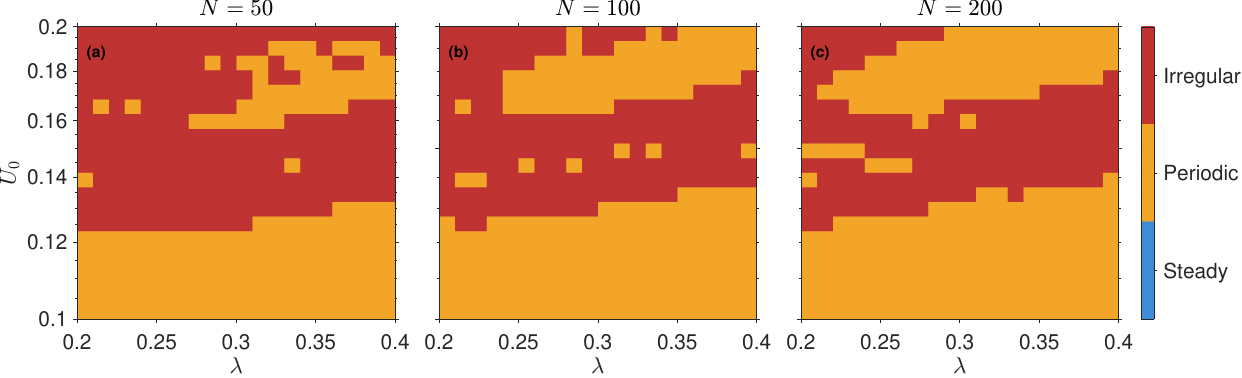}
    \caption{
    \textbf{Sensitivity of the dynamical phase diagram to spatial
    resolution.}
    Dynamical classification over a representative subset of the
    $(\lambda,U_0)$ parameter space using
    (a) $N=50$, (b) $N=100$, and (c) $N=200$ spatial grid points.
    The overall organisation of the steady, periodic and irregular
    regimes is retained with increasing resolution, while the
    detailed boundaries and individual periodic and irregular
    windows show some sensitivity to the spatial discretisation.
    }
    \label{fig:grid_convergence}
\end{figure}

\section{Sensitivity to spatial resolution}
\label{app:grid_convergence}

To assess the sensitivity of the dynamical classification to
spatial discretisation, we repeat the phase-diagram
calculation over a representative region of the
$(\lambda,U_0)$ plane using $N=50$, $100$, and $200$ grid
points. Figure~\ref{fig:grid_convergence} compares the
resulting steady, periodic and irregular regimes.

The broad structure of the phase diagram is consistent across
the three resolutions. In particular, the steady-to-time-dependent
boundary and the occurrence of both periodic and irregular
dynamics at stronger forcing are retained as $N$ is increased.
The detailed boundaries between periodic and irregular states,
however, shift {slightly} with resolution, and individual
periodic or irregular windows can appear or disappear. This
sensitivity is expected in the more complex regime, where
small changes in the spatially resolved dynamics can alter the
long-time attractor and hence its classification. We therefore
interpret the fine-scale alternation between periodic and
irregular regions in Fig.~\ref{fig:phase_diagram} cautiously,
while the overall progression from steady to time-dependent
dynamics is robust to spatial refinement.

\begin{figure*}
    \centering
    \includegraphics[width=\textwidth]{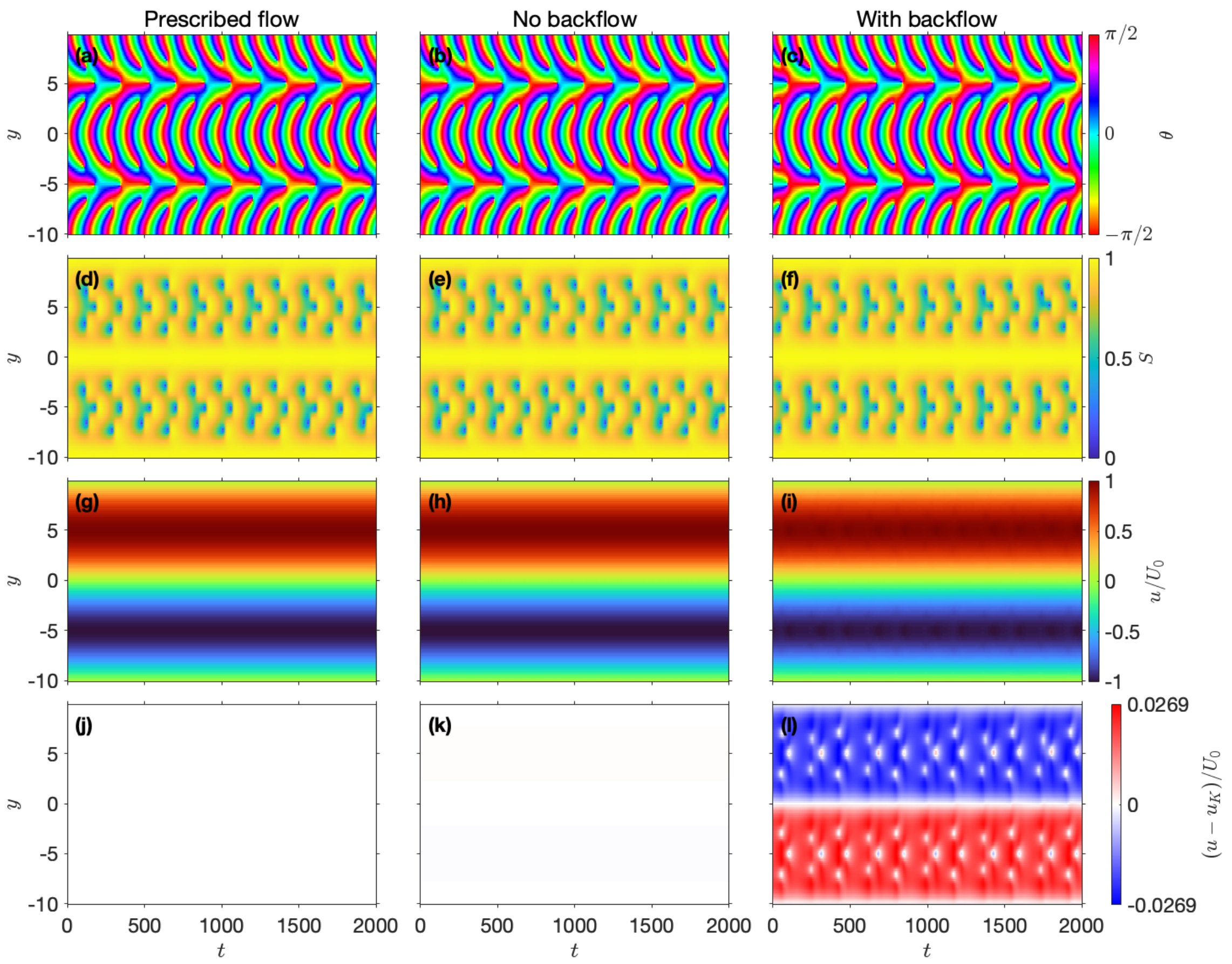}
    \caption{
    \textbf{Effect of velocity dynamics and nematic backflow.}
    Comparison between a prescribed steady Kolmogorov flow
    (left column), a dynamically evolving velocity without
    backflow (middle column), and the fully coupled
    nematohydrodynamic system with backflow (right column).
    From top to bottom, the rows show the director orientation
    $\theta(y,t)$ [(a)--(c)], scalar nematic order $S(y,t)$
    [(d)--(f)], normalized velocity $u(y,t)/U_0$ [(g)--(i)],
    and the normalized deviation from the reference Kolmogorov
    profile, $(u-u_K)/U_0$, with
    $u_K(y)=U_0\sin(k_0y)$ [(j)--(l)].
    For the prescribed-flow model, $u=u_K$ identically [(g),(j)].
    When the velocity evolves without backflow, it remains at
    the Kolmogorov solution to numerical accuracy [(h),(k)],
    and the corresponding nematic dynamics closely reproduce
    those of the prescribed-flow model [(a),(b),(d),(e)].
    Including backflow generates a small but structured,
    time-dependent perturbation of the velocity [(i),(l)],
    correlated with the evolving nematic field.
    The qualitative spatiotemporal dynamics of $\theta$ and $S$
    nevertheless persist [(c),(f)], demonstrating that
    hydrodynamic feedback modifies the dynamics but is not
    required for their emergence.
    Parameters are $U_0=0.2$, $\lambda=0$, $K=0.01$, $\chi=0.1$,
    $\Gamma=1$, $S_{\rm nem}=1$, $\rho=10$, $\eta=4/3$, and
    $L=L_0=20$.
    }
    \label{fig:backflow_comparison}
\end{figure*}

\section{Effect of hydrodynamic feedback}
\label{app:backflow}

The results {described in the main text} were obtained by prescribing the velocity
field and evolving only the nematic order parameter. Here we
test how this reduction compares with the full
nematohydrodynamic problem in which the velocity is allowed to
evolve dynamically. We consider three levels of coupling:
(i) the prescribed Kolmogorov flow used in the main text,
(ii) a dynamically evolving velocity field with nematic
backflow suppressed, and (iii) the fully coupled system with
backflow included.

For the latter two cases, the velocity obeys
\begin{equation}
\rho \partial_t u
=
\eta \partial_y^2 u
+
\rho F \sin(k_0 y)
+
b\,f_{\rm B}.
\label{eq:app_velocity}
\end{equation}
Here $f_B$ denotes the nematic backflow force. For the
$\lambda=0$ case considered in Fig.~\ref{fig:backflow_comparison},
the stress contribution reduces to
\begin{equation}
f_B
=
-2\partial_y
\left(
Q_{xy}H_{xx}-Q_{xx}H_{xy}
\right),
\label{eq:backflow_force}
\end{equation}
where $H_{xx}$ and $H_{xy}$ are given by
Eq.~(\ref{eq:H}). The parameter $b=0$ or $1$ switches the
backflow contribution off or on, respectively. The forcing amplitude is chosen such that, for
$b=0$, the stationary solution is the Kolmogorov profile
\begin{equation}
u_K(y)=U_0\sin(k_0y).
\label{eq:app_kolmogorov}
\end{equation}
The three simulations are initialised with the same nematic
configuration and the same Kolmogorov velocity profile.

Figure~\ref{fig:backflow_comparison} compares the resulting
long-time dynamics. As expected, allowing the velocity to
evolve in the absence of backflow leaves it at the Kolmogorov
solution to numerical accuracy. 

Including backflow produces a small but clearly structured
time-dependent correction to the velocity. To expose this
correction, the bottom row of
Fig.~\ref{fig:backflow_comparison} shows
\begin{equation}
\frac{\delta u(y,t)}{U_0}
=
\frac{u(y,t)-u_K(y)}{U_0}.
\label{eq:app_velocity_deviation}
\end{equation}
While $\delta u$ vanishes for the prescribed and no-backflow
cases, the fully coupled system develops spatiotemporal
variations correlated with the evolving nematic texture.
Nevertheless, the characteristic dynamics of $\theta$ and $S$
remain qualitatively similar to those of the prescribed-flow
model. Thus backflow perturbs and reorganises the nematic dynamics, but
is not required for their existence. This supports the central
conclusion of the main text that the temporal complexity can
originate within the nonlinear dynamics of the nematic order
parameter itself.


\bibliography{complex_nematic}


\begin{thebibliography}{31}
\ifx \bisbn   \undefined \def \bisbn  #1{ISBN #1}\fi
\ifx \binits  \undefined \def \binits#1{#1}\fi
\ifx \bauthor  \undefined \def \bauthor#1{#1}\fi
\ifx \batitle  \undefined \def \batitle#1{#1}\fi
\ifx \bjtitle  \undefined \def \bjtitle#1{#1}\fi
\ifx \bvolume  \undefined \def \bvolume#1{\textbf{#1}}\fi
\ifx \byear  \undefined \def \byear#1{#1}\fi
\ifx \bissue  \undefined \def \bissue#1{#1}\fi
\ifx \bfpage  \undefined \def \bfpage#1{#1}\fi
\ifx \blpage  \undefined \def \blpage #1{#1}\fi
\ifx \burl  \undefined \def \burl#1{\textsf{#1}}\fi
\ifx \doiurl  \undefined \def \doiurl#1{\url{https://doi.org/#1}}\fi
\ifx \betal  \undefined \def \betal{\textit{et al.}}\fi
\ifx \binstitute  \undefined \def \binstitute#1{#1}\fi
\ifx \binstitutionaled  \undefined \def \binstitutionaled#1{#1}\fi
\ifx \bctitle  \undefined \def \bctitle#1{#1}\fi
\ifx \beditor  \undefined \def \beditor#1{#1}\fi
\ifx \bpublisher  \undefined \def \bpublisher#1{#1}\fi
\ifx \bbtitle  \undefined \def \bbtitle#1{#1}\fi
\ifx \bedition  \undefined \def \bedition#1{#1}\fi
\ifx \bseriesno  \undefined \def \bseriesno#1{#1}\fi
\ifx \blocation  \undefined \def \blocation#1{#1}\fi
\ifx \bsertitle  \undefined \def \bsertitle#1{#1}\fi
\ifx \bsnm \undefined \def \bsnm#1{#1}\fi
\ifx \bsuffix \undefined \def \bsuffix#1{#1}\fi
\ifx \bparticle \undefined \def \bparticle#1{#1}\fi
\ifx \barticle \undefined \def \barticle#1{#1}\fi
\bibcommenthead
\ifx \bconfdate \undefined \def \bconfdate #1{#1}\fi
\ifx \botherref \undefined \def \botherref #1{#1}\fi
\ifx \url \undefined \def \url#1{\textsf{#1}}\fi
\ifx \bchapter \undefined \def \bchapter#1{#1}\fi
\ifx \bbook \undefined \def \bbook#1{#1}\fi
\ifx \bcomment \undefined \def \bcomment#1{#1}\fi
\ifx \oauthor \undefined \def \oauthor#1{#1}\fi
\ifx \citeauthoryear \undefined \def \citeauthoryear#1{#1}\fi
\ifx \endbibitem  \undefined \def \endbibitem {}\fi
\ifx \bconflocation  \undefined \def \bconflocation#1{#1}\fi
\ifx \arxivurl  \undefined \def \arxivurl#1{\textsf{#1}}\fi
\csname PreBibitemsHook\endcsname

\bibitem[\protect\citeauthoryear{de~Gennes and Prost}{1993}]{de1993physics}
\begin{bbook}
\bauthor{\bsnm{Gennes}, \binits{P.-G.}},
\bauthor{\bsnm{Prost}, \binits{J.}}:
\bbtitle{The Physics of Liquid Crystals},
\bedition{2}nd edn.
\bpublisher{Clarendon Press},
\blocation{Oxford}
(\byear{1993}).
\doiurl{10.1093/oso/9780198520245.001.0001}
\end{bbook}
\endbibitem

\bibitem[\protect\citeauthoryear{Lewis et~al.}{2014}]{lewis2014colloidal}
\begin{barticle}
\bauthor{\bsnm{Lewis}, \binits{A.H.}},
\bauthor{\bsnm{Garlea}, \binits{I.}},
\bauthor{\bsnm{Alvarado}, \binits{J.}},
\bauthor{\bsnm{Dammone}, \binits{O.J.}},
\bauthor{\bsnm{Howell}, \binits{P.D.}},
\bauthor{\bsnm{Majumdar}, \binits{A.}},
\bauthor{\bsnm{Mulder}, \binits{B.M.}},
\bauthor{\bsnm{Lettinga}, \binits{M.}},
\bauthor{\bsnm{Koenderink}, \binits{G.H.}},
\bauthor{\bsnm{Aarts}, \binits{D.G.}}:
\batitle{Colloidal liquid crystals in rectangular confinement: theory and experiment}.
\bjtitle{Soft Matter}
\bvolume{10}(\bissue{39}),
\bfpage{7865}--\blpage{7873}
(\byear{2014})
\doiurl{10.1039/C4SM01123F}
\end{barticle}
\endbibitem

\bibitem[\protect\citeauthoryear{Doostmohammadi and Ladoux}{2022}]{doostmohammadi2022physics}
\begin{barticle}
\bauthor{\bsnm{Doostmohammadi}, \binits{A.}},
\bauthor{\bsnm{Ladoux}, \binits{B.}}:
\batitle{Physics of liquid crystals in cell biology}.
\bjtitle{Trends in cell biology}
\bvolume{32}(\bissue{2}),
\bfpage{140}--\blpage{150}
(\byear{2022})
\doiurl{10.1016/j.tcb.2021.09.012}
\end{barticle}
\endbibitem

\bibitem[\protect\citeauthoryear{Maroudas-Sacks et~al.}{2021}]{maroudas2021topological}
\begin{barticle}
\bauthor{\bsnm{Maroudas-Sacks}, \binits{Y.}},
\bauthor{\bsnm{Garion}, \binits{L.}},
\bauthor{\bsnm{Shani-Zerbib}, \binits{L.}},
\bauthor{\bsnm{Livshits}, \binits{A.}},
\bauthor{\bsnm{Braun}, \binits{E.}},
\bauthor{\bsnm{Keren}, \binits{K.}}:
\batitle{Topological defects in the nematic order of actin fibres as organization centres of {H}ydra morphogenesis}.
\bjtitle{Nature Physics}
\bvolume{17}(\bissue{2}),
\bfpage{251}--\blpage{259}
(\byear{2021})
\doiurl{10.1038/s41567-020-01083-1}
\end{barticle}
\endbibitem

\bibitem[\protect\citeauthoryear{Saw et~al.}{2017}]{saw2017topological}
\begin{barticle}
\bauthor{\bsnm{Saw}, \binits{T.B.}},
\bauthor{\bsnm{Doostmohammadi}, \binits{A.}},
\bauthor{\bsnm{Nier}, \binits{V.}},
\bauthor{\bsnm{Kocgozlu}, \binits{L.}},
\bauthor{\bsnm{Thampi}, \binits{S.}},
\bauthor{\bsnm{Toyama}, \binits{Y.}},
\bauthor{\bsnm{Marcq}, \binits{P.}},
\bauthor{\bsnm{Lim}, \binits{C.T.}},
\bauthor{\bsnm{Yeomans}, \binits{J.M.}},
\bauthor{\bsnm{Ladoux}, \binits{B.}}:
\batitle{Topological defects in epithelia govern cell death and extrusion}.
\bjtitle{Nature}
\bvolume{544}(\bissue{7649}),
\bfpage{212}--\blpage{216}
(\byear{2017})
\doiurl{10.1038/nature21718}
\end{barticle}
\endbibitem

\bibitem[\protect\citeauthoryear{Shankar et~al.}{2022}]{Shankar2022}
\begin{barticle}
\bauthor{\bsnm{Shankar}, \binits{S.}},
\bauthor{\bsnm{Souslov}, \binits{A.}},
\bauthor{\bsnm{Bowick}, \binits{M.J.}},
\bauthor{\bsnm{Marchetti}, \binits{M.C.}},
\bauthor{\bsnm{Vitelli}, \binits{V.}}:
\batitle{Topological active matter}.
\bjtitle{Nature Reviews Physics}
\bvolume{4}(\bissue{6}),
\bfpage{380}--\blpage{398}
(\byear{2022})
\doiurl{10.1038/s42254-022-00445-3}
\end{barticle}
\endbibitem

\bibitem[\protect\citeauthoryear{Giomi}{2015}]{giomi2015geometry}
\begin{barticle}
\bauthor{\bsnm{Giomi}, \binits{L.}}:
\batitle{Geometry and topology of turbulence in active nematics}.
\bjtitle{Physical Review X}
\bvolume{5}(\bissue{3}),
\bfpage{031003}
(\byear{2015})
\doiurl{10.1103/PhysRevX.5.031003}
\end{barticle}
\endbibitem

\bibitem[\protect\citeauthoryear{Doostmohammadi et~al.}{2018}]{Doostmohammadi2018_ActiveNematics}
\begin{barticle}
\bauthor{\bsnm{Doostmohammadi}, \binits{A.}},
\bauthor{\bsnm{Ign{\'e}s-Mullol}, \binits{J.}},
\bauthor{\bsnm{Yeomans}, \binits{J.M.}},
\bauthor{\bsnm{Sagu{\'e}s}, \binits{F.}}:
\batitle{Active nematics}.
\bjtitle{Nature Communications}
\bvolume{9}(\bissue{1}),
\bfpage{3246}
(\byear{2018})
\doiurl{10.1038/s41467-018-05666-8}
\end{barticle}
\endbibitem

\bibitem[\protect\citeauthoryear{Alert et~al.}{2022}]{alert2022active}
\begin{barticle}
\bauthor{\bsnm{Alert}, \binits{R.}},
\bauthor{\bsnm{Casademunt}, \binits{J.}},
\bauthor{\bsnm{Joanny}, \binits{J.-F.}}:
\batitle{Active turbulence}.
\bjtitle{Annual Review of Condensed Matter Physics}
\bvolume{13}(\bissue{1}),
\bfpage{143}--\blpage{170}
(\byear{2022})
\doiurl{10.1146/annurev-conmatphys-082321-035957}
\end{barticle}
\endbibitem

\bibitem[\protect\citeauthoryear{Thampi et~al.}{2016}]{Thampi2016}
\begin{barticle}
\bauthor{\bsnm{Thampi}, \binits{S.P.}},
\bauthor{\bsnm{Golestanian}, \binits{R.}},
\bauthor{\bsnm{Yeomans}, \binits{J.M.}}:
\batitle{Vorticity, defects and correlations in active turbulence}.
\bjtitle{Philosophical Transactions of the Royal Society A: Mathematical, Physical and Engineering Sciences}
\bvolume{374}(\bissue{2080}),
\bfpage{20150130}
(\byear{2016})
\doiurl{10.1098/rsta.2015.0130}
\end{barticle}
\endbibitem

\bibitem[\protect\citeauthoryear{Dubois-Violette et~al.}{1971}]{dubois1971hydrodynamic}
\begin{barticle}
\bauthor{\bsnm{Dubois-Violette}, \binits{E.}},
\bauthor{\bsnm{De~Gennes}, \binits{P.}},
\bauthor{\bsnm{Parodi}, \binits{O.}}:
\batitle{Hydrodynamic instabilities of nematic liquid crystals under {AC} electric fields}.
\bjtitle{Journal de Physique}
\bvolume{32}(\bissue{4}),
\bfpage{305}--\blpage{317}
(\byear{1971})
\doiurl{10.1051/jphys:01971003204030500}
\end{barticle}
\endbibitem

\bibitem[\protect\citeauthoryear{Manneville}{1981}]{manneville1981transition}
\begin{barticle}
\bauthor{\bsnm{Manneville}, \binits{P.}}:
\batitle{The transition to turbulence in nematic liquid crystals}.
\bjtitle{Molecular crystals and liquid crystals}
\bvolume{70}(\bissue{1}),
\bfpage{223}--\blpage{250}
(\byear{1981})
\doiurl{10.1080/00268948108073591}
\end{barticle}
\endbibitem

\bibitem[\protect\citeauthoryear{Weiss and Ahlers}{2013}]{weiss2013nematic}
\begin{barticle}
\bauthor{\bsnm{Weiss}, \binits{S.}},
\bauthor{\bsnm{Ahlers}, \binits{G.}}:
\batitle{Nematic--isotropic phase transition in turbulent thermal convection}.
\bjtitle{Journal of fluid mechanics}
\bvolume{737},
\bfpage{308}--\blpage{328}
(\byear{2013})
\doiurl{10.1017/jfm.2013.563}
\end{barticle}
\endbibitem

\bibitem[\protect\citeauthoryear{Jewell et~al.}{2009}]{Jewell2009}
\begin{barticle}
\bauthor{\bsnm{Jewell}, \binits{S.A.}},
\bauthor{\bsnm{Cornford}, \binits{S.L.}},
\bauthor{\bsnm{Yang}, \binits{F.}},
\bauthor{\bsnm{Cann}, \binits{P.S.}},
\bauthor{\bsnm{Sambles}, \binits{J.R.}}:
\batitle{Flow-driven transition and associated velocity profiles in a nematic liquid‐crystal cell}.
\bjtitle{Physical Review E}
\bvolume{80}(\bissue{4}),
\bfpage{041706}
(\byear{2009})
\doiurl{10.1103/PhysRevE.80.041706}
\end{barticle}
\endbibitem

\bibitem[\protect\citeauthoryear{Sengupta et~al.}{2013}]{Sengupta2013}
\begin{barticle}
\bauthor{\bsnm{Sengupta}, \binits{A.}},
\bauthor{\bsnm{Tkalec}, \binits{U.}},
\bauthor{\bsnm{Ravnik}, \binits{M.}},
\bauthor{\bsnm{Yeomans}, \binits{J.M.}},
\bauthor{\bsnm{Bahr}, \binits{C.}},
\bauthor{\bsnm{Herminghaus}, \binits{S.}}:
\batitle{Liquid crystal microfluidics for tunable flow shaping}.
\bjtitle{Physical Review Letters}
\bvolume{110},
\bfpage{048303}
(\byear{2013})
\doiurl{10.1103/PhysRevLett.110.048303}
\end{barticle}
\endbibitem

\bibitem[\protect\citeauthoryear{Aplinc et~al.}{2016}]{aplinc2016porous}
\begin{barticle}
\bauthor{\bsnm{Aplinc}, \binits{J.}},
\bauthor{\bsnm{Morris}, \binits{S.}},
\bauthor{\bsnm{Ravnik}, \binits{M.}}:
\batitle{Porous nematic microfluidics for generation of umbilic defects and umbilic defect lattices}.
\bjtitle{Physical Review Fluids}
\bvolume{1}(\bissue{2}),
\bfpage{023303}
(\byear{2016})
\doiurl{10.1103/PhysRevFluids.1.023303}
\end{barticle}
\endbibitem

\bibitem[\protect\citeauthoryear{Giomi et~al.}{2017}]{giomi2017cross}
\begin{barticle}
\bauthor{\bsnm{Giomi}, \binits{L.}},
\bauthor{\bsnm{Kos}, \binits{{\v{Z}}.}},
\bauthor{\bsnm{Ravnik}, \binits{M.}},
\bauthor{\bsnm{Sengupta}, \binits{A.}}:
\batitle{Cross-talk between topological defects in different fields revealed by nematic microfluidics}.
\bjtitle{Proceedings of the National Academy of Sciences}
\bvolume{114}(\bissue{29}),
\bfpage{5771}--\blpage{5777}
(\byear{2017})
\doiurl{10.1073/pnas.1702777114}
\end{barticle}
\endbibitem

\bibitem[\protect\citeauthoryear{{\v C}opar et~al.}{2020}]{Copar2020-qu}
\begin{barticle}
\bauthor{\bsnm{{\v C}opar}, \binits{S.}},
\bauthor{\bsnm{Kos}, \binits{{\v Z}.}},
\bauthor{\bsnm{Emer{\v s}i{\v c}}, \binits{T.}},
\bauthor{\bsnm{Tkalec}, \binits{U.}}:
\batitle{Microfluidic control over topological states in channel-confined nematic flows}.
\bjtitle{Nature Communications}
\bvolume{11}(\bissue{1}),
\bfpage{59}
(\byear{2020})
\doiurl{10.1038/s41467-019-13789-9}
\end{barticle}
\endbibitem

\bibitem[\protect\citeauthoryear{Mur et~al.}{2022}]{mur2022continuous}
\begin{barticle}
\bauthor{\bsnm{Mur}, \binits{M.}},
\bauthor{\bsnm{Kos}, \binits{{\v{Z}}.}},
\bauthor{\bsnm{Ravnik}, \binits{M.}},
\bauthor{\bsnm{Mu{\v{s}}evi{\v{c}}}, \binits{I.}}:
\batitle{Continuous generation of topological defects in a passively driven nematic liquid crystal}.
\bjtitle{Nature Communications}
\bvolume{13}(\bissue{1}),
\bfpage{6855}
(\byear{2022})
\doiurl{10.1038/s41467-022-34384-5}
\end{barticle}
\endbibitem

\bibitem[\protect\citeauthoryear{Valani et~al.}{2026}]{Valani_Thampi_Yeomans_2026}
\begin{barticle}
\bauthor{\bsnm{Valani}, \binits{R.N.}},
\bauthor{\bsnm{Thampi}, \binits{S.P.}},
\bauthor{\bsnm{Yeomans}, \binits{J.M.}}:
\batitle{From equilibrium multistability to spatiotemporal chaos in channel flows of nematic fluids}.
\bjtitle{Journal of Fluid Mechanics}
\bvolume{1038},
\bfpage{51}
(\byear{2026})
\doiurl{10.1017/jfm.2026.11800}
\end{barticle}
\endbibitem

\bibitem[\protect\citeauthoryear{Beris and Edwards}{1994}]{Beris1994}
\begin{bbook}
\bauthor{\bsnm{Beris}, \binits{A.N.}},
\bauthor{\bsnm{Edwards}, \binits{B.J.}}:
\bbtitle{Thermodynamics of Flowing Systems: With Internal Microstructure}.
\bpublisher{Oxford University Press},
\blocation{Oxford}
(\byear{1994}).
\doiurl{10.1093/oso/9780195076943.001.0001}
\end{bbook}
\endbibitem

\bibitem[\protect\citeauthoryear{Híjar and Majumdar}{2024}]{D3SM01684F}
\begin{barticle}
\bauthor{\bsnm{Híjar}, \binits{H.}},
\bauthor{\bsnm{Majumdar}, \binits{A.}}:
\batitle{Particle-based and continuum models for confined nematics in two dimensions}.
\bjtitle{Soft Matter}
\bvolume{20}(\bissue{18}),
\bfpage{3755}--\blpage{3770}
(\byear{2024})
\doiurl{10.1039/D3SM01684F}
\end{barticle}
\endbibitem

\bibitem[\protect\citeauthoryear{Grosso et~al.}{2001}]{Grosso2001}
\begin{barticle}
\bauthor{\bsnm{Grosso}, \binits{M.}},
\bauthor{\bsnm{Keunings}, \binits{R.}},
\bauthor{\bsnm{Crescitelli}, \binits{S.}},
\bauthor{\bsnm{Maffettone}, \binits{P.L.}}:
\batitle{Prediction of chaotic dynamics in sheared liquid crystalline polymers}.
\bjtitle{Physical Review Letters}
\bvolume{86}(\bissue{14}),
\bfpage{3184}--\blpage{3187}
(\byear{2001})
\doiurl{10.1103/PhysRevLett.86.3184}
\end{barticle}
\endbibitem

\bibitem[\protect\citeauthoryear{Forest et~al.}{2004}]{Forest2004}
\begin{barticle}
\bauthor{\bsnm{Forest}, \binits{M.G.}},
\bauthor{\bsnm{Wang}, \binits{Q.}},
\bauthor{\bsnm{Zhou}, \binits{R.}}:
\batitle{The flow-phase diagram of {Doi--Hess} theory for sheared nematic polymers. {II}. finite shear rates}.
\bjtitle{Rheologica Acta}
\bvolume{44}(\bissue{1}),
\bfpage{80}--\blpage{93}
(\byear{2004})
\doiurl{10.1007/s00397-004-0380-9}
\end{barticle}
\endbibitem

\bibitem[\protect\citeauthoryear{Rien{\"a}cker et~al.}{2002}]{RienaeckerKroegerHess2002}
\begin{barticle}
\bauthor{\bsnm{Rien{\"a}cker}, \binits{G.}},
\bauthor{\bsnm{Kr{\"o}ger}, \binits{M.}},
\bauthor{\bsnm{Hess}, \binits{S.}}:
\batitle{Chaotic orientational behavior of a nematic liquid crystal subjected to a steady shear flow}.
\bjtitle{Physical Review E}
\bvolume{66},
\bfpage{040702}
(\byear{2002})
\doiurl{10.1103/PhysRevE.66.040702}
\end{barticle}
\endbibitem

\bibitem[\protect\citeauthoryear{Chakrabarti et~al.}{2004}]{Chakrabarti2004}
\begin{barticle}
\bauthor{\bsnm{Chakrabarti}, \binits{B.}},
\bauthor{\bsnm{Das}, \binits{M.}},
\bauthor{\bsnm{Dasgupta}, \binits{C.}},
\bauthor{\bsnm{Ramaswamy}, \binits{S.}},
\bauthor{\bsnm{Sood}, \binits{A.K.}}:
\batitle{Spatiotemporal rheochaos in nematic hydrodynamics}.
\bjtitle{Physical Review Letters}
\bvolume{92},
\bfpage{055501}
(\byear{2004})
\doiurl{10.1103/PhysRevLett.92.055501}
\end{barticle}
\endbibitem

\bibitem[\protect\citeauthoryear{Forest et~al.}{2005}]{Forest2005}
\begin{barticle}
\bauthor{\bsnm{Forest}, \binits{M.G.}},
\bauthor{\bsnm{Zhou}, \binits{R.}},
\bauthor{\bsnm{Wang}, \binits{Q.}}:
\batitle{Kinetic structure simulations of nematic polymers in plane {Couette} cells. {II}: In-plane structure transitions}.
\bjtitle{Multiscale Modeling \& Simulation}
\bvolume{4}(\bissue{4}),
\bfpage{1280}--\blpage{1304}
(\byear{2005})
\doiurl{10.1137/040618187}
\end{barticle}
\endbibitem

\bibitem[\protect\citeauthoryear{Forest et~al.}{2008}]{Forest2008}
\begin{barticle}
\bauthor{\bsnm{Forest}, \binits{M.G.}},
\bauthor{\bsnm{Heidenreich}, \binits{S.}},
\bauthor{\bsnm{Hess}, \binits{S.}},
\bauthor{\bsnm{Yang}, \binits{X.}},
\bauthor{\bsnm{Zhou}, \binits{R.}}:
\batitle{Robustness of pulsating jet-like layers in sheared nano-rod dispersions}.
\bjtitle{Journal of Non-Newtonian Fluid Mechanics}
\bvolume{155}(\bissue{3}),
\bfpage{130}--\blpage{145}
(\byear{2008})
\doiurl{10.1016/j.jnnfm.2008.06.003}
\end{barticle}
\endbibitem

\bibitem[\protect\citeauthoryear{Cross and Hohenberg}{1993}]{CrossHohenberg1993}
\begin{barticle}
\bauthor{\bsnm{Cross}, \binits{M.C.}},
\bauthor{\bsnm{Hohenberg}, \binits{P.C.}}:
\batitle{Pattern formation outside of equilibrium}.
\bjtitle{Reviews of Modern Physics}
\bvolume{65},
\bfpage{851}--\blpage{1112}
(\byear{1993})
\doiurl{10.1103/RevModPhys.65.851}
\end{barticle}
\endbibitem

\bibitem[\protect\citeauthoryear{Cross and Greenside}{2009}]{CrossGreenside2009}
\begin{bbook}
\bauthor{\bsnm{Cross}, \binits{M.C.}},
\bauthor{\bsnm{Greenside}, \binits{H.}}:
\bbtitle{Pattern Formation and Dynamics in Nonequilibrium Systems}.
\bpublisher{Cambridge University Press},
\blocation{Cambridge}
(\byear{2009}).
\doiurl{10.1017/CBO9780511627200}
\end{bbook}
\endbibitem

\bibitem[\protect\citeauthoryear{Strogatz}{2015}]{Strogatz2015_NonlinearDynamicsChaos}
\begin{bbook}
\bauthor{\bsnm{Strogatz}, \binits{S.H.}}:
\bbtitle{Nonlinear Dynamics and Chaos: With Applications to Physics, Biology, Chemistry, and Engineering},
\bedition{2}nd edn.
\bpublisher{Westview Press},
\blocation{Boulder, CO}
(\byear{2015})
\end{bbook}
\endbibitem

\end{thebibliography}

\end{document}